\documentclass[11pt]{article}
\title
{A Strategy for Maker in the Clique Game \\ which Helps to Tackle some Open Problems by Beck}
\author{Heidi Gebauer
\thanks{Institute of
Theoretical Computer Science, ETH Zurich, CH-8092 Switzerland. Email:
gebauerh@inf.ethz.ch. } 
}

\date{}
\usepackage{amsmath, amsfonts}
\usepackage{amsthm}
\usepackage{subfigure}
\usepackage[dvips]{graphicx}

\begin{document}
\bibliographystyle{plain}
\maketitle
% Theorems, etc.
\newtheorem{theo}{Theorem} [section]
\newtheorem{defi}[theo]{Definition}
\newtheorem{lemm}[theo]{Lemma}                                                                                                                                                                                                                                                                                                                                                                                                                                                                                                                   
\newtheorem{obse}[theo]{Observation}
\newtheorem{prop}[theo]{Proposition}
\newtheorem{coro}[theo]{Corollary}
\newtheorem{rem}[theo]{Remark}
\newtheorem{opprob}[theo]{Open Problem}

%\newtheorem{theorem}{Theorem}
%\newtheorem{claim}{Claim}
%\newtheorem{lemma}{Lemma}
%\newtheorem{propos}{Proposition}
%\newtheorem{conjecture}{Conjecture}
%\newtheorem{problem}{Problem}
%\newtheorem{corol}{Corollary}
%\newcommand{\Proof}{\noindent{\bf Proof.}\ \ }
%\newcommand{\Remarks}{\noindent{\bf Remarks:}\ \ }
%%Probability
\newcommand{\whp}{{\bf whp}}
\newcommand{\prob}{probability}
\newcommand{\rn}{random}
\newcommand{\rv}{random variable}
%Hypergraphs
\newcommand{\hpg}{hypergraph}
\newcommand{\hpgs}{hypergraphs}
\newcommand{\subhpg}{subhypergraph}
\newcommand{\subhpgs}{subhypergraphs}
%Letters
\newcommand{\bH}{{\bf H}}
\newcommand{\cH}{{\cal H}}
\newcommand{\cT}{{\cal T}}
\newcommand{\cF}{{\cal F}}
\newcommand{\cD}{{\cal D}}
\newcommand{\cC}{{\cal C}}

\newcommand{\ideg}{\mathsf {ideg}}
\newcommand{\lv}{\mathsf {lv}}
\newcommand{\nga}{n_{\text{game}}}
\newcommand{\avdeg}{\overline{\deg}}
\newcommand{\ed}{e_{\text{double}}}

%commands min deg c
\newcommand{\degb}{\deg_{B}}
\newcommand{\degm}{\deg_{M}}

\newcommand{\avd}{\bar{D}}

\newcommand{\remain}{\mathsf {rem}}

\begin{abstract}
We study Maker/Breaker games on the edges of the 
complete graph, as introduced by Chv\'atal and Erd\H os.
We show that in the $(m:b)$ clique game played on $K_{N}$, the complete graph on $N$ vertices, Maker can achieve a 
$K_{q}$ for $q =  \left(\frac{m}{\log_{2}(b + 1)} - o(1)\right) \cdot \log N$, which partially solves an open problem by Beck.
Moreover, we show that in the (1:1) clique game played on $K_{N}$ for a sufficiently large $N$, Maker can achieve a $K_{q}$ in only $O(2^{\frac{2q}{3}})$ moves, which improves the previous best bound and answers a question of Beck.
Finally we consider the so called \emph{tournament game}. 
A \emph{tournament} is a directed graph where every pair of vertices is connected by a single directed edge. The tournament game is played on $K_{N}$.
At the beginning Breaker fixes an arbitrary tournament $T_{q}$ on $q$ vertices. Maker and Breaker then alternately take turns at claiming one unclaimed edge $e$ and selecting one of the two possible orientations. Maker wins if his graph contains a copy of the goal tournament $T_{q}$; otherwise Breaker wins.
We show that Maker wins the tournament game on $K_{N}$ with $q = (1 - o(1)) \log_{2} N$ which supports the random graph intuition: the threshold for $q$ is {\em asymptotically the same} for the game
played by two ``clever'' players and the game played by two ``random''
players.

This last result solves an open problem of Beck which he included in his list of the \emph{seven most humiliating open problems}. 

\end{abstract}

\section{Introduction}

In this paper we study games played by two opponents on edges of the complete graph $K_N$ on $N$ vertices. The two players alternately take turns at claiming some number of unclaimed edges until all edges are claimed. 
One of the players, called Maker, aims to create such a graph which possesses some fixed property $P$. The other player,
called Breaker, tries to prevent Maker from achieving his goal: Breaker wins if, after all ${n\choose 2}$ edges were claimed, Maker's graph does {\em not} posses $P$. A widely studied game of this kind is the \emph{$q$-clique game} where $P = K_{q}$, the clique on $q$ vertices.
An immediate question is how large $q$ can be (in terms of $N$) such that Maker can achieve a $K_{q}$ in the game on $K_{N}$.
Amazingly, for the ordinary $(1:1)$ game, i.e., the game where Maker and Breaker each take one edge per turn, the 
exact solution to this question is known!
%question has been completely solved. 
Let $f(N) := 2 \log N - 2 \log \log N + 2 \log e - 3$.
(Throughout this paper $\log$ stands for the binary logarithm.)
\begin{theo} \label{theo:exactsolforcliquegame}
(Beck, \cite{BeckTTT})
If $q \leq \lfloor f(N) \rfloor$ then Maker has a winning strategy in the (1:1) $q$-clique game.
If $q \geq \lceil f(N) \rceil$ then Breaker has a winning strategy.
\end{theo}
\noindent
For the \emph{biased} variant of the $q$-clique game, where Maker claims, say, $m$ edges per move and Breaker claims, say, $b$ edges per move, however, not so much is now. 
We denote this game by \emph{$(m:b)$ clique game}
and let $f(N, m, b)$ denote the largest $q$ such that Maker can achieve a $K_{q}$ in the $(m:b)$ clique game on $K_{N}$. Let
$g(N,m,b) =  $
\newline
\newline
$\left\{ 
\begin{array}{ll}
\left\lfloor \frac{2}{\log(m + b) - \log m} \cdot \log N - 2 \log_{c} \log_{c} N + 2 \log_{c} e - 2 \log_{c} 2 - 1 + 
\frac{2 \log c}{\log c_{0}} + o(1) \right\rfloor, & \text{if } m > b \\
& \\
\left\lfloor \frac{2}{\log(m + b) - \log m} \cdot \log N - 2 \log_{c} \log_{c} N + 2 \log_{c} e - 2 \log_{c} 2 - 1 + o(1) \right\rfloor, & \text{if } m \leq b 
\end{array}
\right.
$
\newline
\newline
where $c = \frac{m + b}{m}$ and $c_{0} = \frac{m}{m-b}$. 
\newline
\newline
Beck poses the following open problem. 

\begin{opprob}  (Open Problem 30.2, \cite{BeckTTT})  \label{opprob:BiasedCliqueMakerBreakerBeck} 
Is it true that 
$f(N,m,b) = g(N,m,b)$?
% $f(N,m,b) =  $
% \newline
% \newline
% $\left\{ 
% \begin{array}{ll}
% \left\lfloor \frac{2}{\log(m + b) - \log m} \cdot \log N - 2 \log_{c} \log_{c} N + 2 \log_{c} e - 2 \log_{c} 2 - 1 + 
% \frac{2 \log c}{\log c_{0}} + o(1) \right\rfloor, & \text{if } m > b \\
% & \\
% \left\lfloor \frac{2}{\log(m + b) - \log m} \cdot \log N - 2 \log_{c} \log_{c} N + 2 \log_{c} e - 2 \log_{c} 2 - 1 + o(1) \right\rfloor, & \text{if } m \leq b 
% \end{array}
% \right.
% $
% \newline
% \newline
% where $c = \frac{m + b}{m}$ and $c_{0} = \frac{m}{m-b}$ ?
\end{opprob}  
\noindent
We will show that this does not hold in full generality by proving the following.

\begin{theo} \label{theo:Makerbuildscliquebiased}
In the $(m:b)$ clique game played on $K_{N}$ Maker has a strategy to achieve a $K_{q}$ with 
$q = \left(\frac{m}{\log(b + 1)} - o(1)\right) \cdot \log N$
\end{theo}
\noindent
If $m$ and $b$ are close to each other (say $\frac{b}{2} \leq m \leq 2b$) and large enough then by Theorem \ref{theo:Makerbuildscliquebiased},
\newline
$f(N,m,b) \geq \left(\frac{m}{\log(2m + 1)} - o(1)\right) \log N
> \left(\frac{2}{\log (\frac{3}{2})} + o(1)\right) \log N \geq g(N,m,b)$.
\newline
In particular, 
%if $m = b$ then in the $(m:b)$ clique game on $K_{N}$ Maker can achieve a $K_{q}$ with$q =  (\frac{m}{\log(m + 1)} - o(1)) \cdot \log N$$ which is larger than $g(N,m,m) = (2 + o(1)) \log N$ for $m \geq 6$.
$f(N, m, m) \geq \left(\frac{m}{\log (m + 1)} - o(1) \right) \log N  > (2 + o(1)) \log N  = g(N,m,m)$ for $m \geq 6$. 
This connects to the following open problem by Beck.

\begin{opprob}  (Open Problem 31.1, \cite{BeckTTT}) \label{opprob:cliquewith2each}
\begin{itemize}
 \item[(a)] Is it true that in the (2:2) clique game on $K_{N}$ Maker has a strategy to achieve a $K_{q}$ for $q = 2 \log N - 2 \log \log N + O(1)$?
 \item[(b)] Is it true that in the (2:2) clique game on $K_{N}$ Breaker can prevent Maker from achieving a $K_{q}$ for $q = 2 \log N - 2 \log \log N + O(1)$?
\end{itemize}
\end{opprob}
\noindent
Open Problem \ref{opprob:cliquewith2each} is still unsolved but Theorem \ref{theo:Makerbuildscliquebiased} (for $m = b \geq 6$) points out that it is not implausible that the answer to (b) is no.
%suggests that the answer for (a) is yes whereas the answer for (b) is no.

Theorem \ref{theo:exactsolforcliquegame}, Open Problem  \ref{opprob:BiasedCliqueMakerBreakerBeck}, Theorem \ref{theo:Makerbuildscliquebiased} and Open Problem \ref{opprob:cliquewith2each} are about the issue of building a $q$-clique on a complete graph containing as few vertices as possible.
Another issue is to build a clique \emph{fast}.

\begin{opprob} (Open Problem 25.1, \cite{BeckTTT})
Playing the (1:1) clique game on the infinite complete graph $K_{\infty}$ (or at least a ``very large'' finite $K_{N}$), how long does it take for Maker to build a $K_{q}$?
\end{opprob}
\noindent
Let $s(q)$ denote the minimum number of moves Maker needs to achieve a $K_{q}$.
Theorem \ref{theo:exactsolforcliquegame} directly implies the following.
\begin{coro} \label{coro:cliquegameoneoneMakerq}
Maker can build a $K_{q}$ on the graph $K_{N}$ with $N \geq q \cdot 2^{\frac{q}{2}}(1 + o(1))$.
\end{coro}
\noindent
Hence $s(q) \leq  \frac{1}{2} \binom{N}{2} \leq (1 + o(1)) q^{2} 2^{q}$. The best known bound to our knowledge is $s(q) \leq 2^{q + 2}$ which has been discovered by Beck \cite{BeckRG} and, independently, by Peke\v{c} and Tuza.
From the other side, Breaker can prevent Maker from building a $K_{q}$ in $2^{\frac{q}{2}}$ moves \cite{BeckTTT}, thus $s(q) \geq 2^{\frac{q}{2}}$.

Beck asks whether the bound $O(2^{q})$ can be improved and, if yes, whether $s(q)$ is closer to the upper bound of (roughly) $2^{q}$ or to the lower bound of $2^{\frac{q}{2}}$. 
We will show that $s(q) \leq O\left(2^{\frac{2q}{3}}\right)$, which means that $s(q)$ actually is closer to the lower bound.

\begin{theo} \label{theo:Makerbuildsfastclique}
Let $q,r$ be integers such that $q$ is sufficiently large and let $N \geq q^{5} \cdot 2^{q} \cdot r$.
In the $(1:1)$ game on $K_{N}$ Maker can in $q^{5} \cdot 2^{q} \cdot r$ moves
achieve that for some $\{v_{1}, \ldots, v_{q}\} \cup \{w_{1}, \ldots, w_{r}\} \subseteq V(G)$, (i) every edge $(v_{i}, u)$ with $i \in \{1, \ldots, q\}$ and $u \in \{v_{1}, \ldots, v_{i-1}, v_{i+1}, \ldots, v_{q}\} \cup \{w_{1}, \ldots, w_{r}\}$ belongs to Maker's graph, and (ii) for every pair $i,j \in \binom{\{1, \ldots, r\}}{2}$, the edge $(w_{i}, w_{j})$ has not been claimed by Breaker.
\end{theo}
\noindent
We can now combine Theorem \ref{theo:Makerbuildsfastclique} and Theorem \ref{theo:exactsolforcliquegame}: First we apply Theorem \ref{theo:Makerbuildsfastclique}, which allows Maker to obtain in his graph a $q$-clique $C$ and a vertex set $\{w_{1}, \ldots, w_{r}\}$ where every $w_{i}$ is connected to every vertex of $C$. Then we apply Theorem \ref{theo:exactsolforcliquegame}, which lets Maker build a $K_{2 \log r - 2 \log \log r - 2}$ on the subgraph induced by $\{w_{1}, \ldots, w_{r}\}$. Hence altogether Maker can achieve a $K_{q + 2 \log r - 3 \log \log r}$ in $q^{5} \cdot 2^{q} \cdot r + \binom{r}{2}$ moves. If we replace $q$ with $\frac{q}{3}$ and $r$ with $q^{2}2^{\frac{q}{3}}$ we obtain the following.

\begin{coro}
For a large enough $N$
Maker can build a $K_{q}$ in the game on $K_{N}$ in $2q^{7} \cdot 2^{\frac{2q}{3}}$ moves
\end{coro}
\noindent

Another variation of the clique game is the so called \emph{tournament game}. A \emph{tournament} is a directed graph where every pair of vertices is connected by a single directed edge. The tournament game is played on $K_{N}$.
At the beginning Breaker fixes an arbitrary tournament $T_{q}$ on $q$ vertices. Maker and Breaker then alternately take turns at claiming one unclaimed edge $e$ and selecting one of the two possible orientations. Maker wins if his graph contains a copy of the goal tournament $T_{q}$; otherwise Breaker wins. Beck \cite{BeckTTT} showed that Maker has a winning strategy for $q = (\frac{1}{2} - o(1)) \log N$. Actually, he even proved the stronger statement that Maker has a strategy to achieve that his graph contains a copy of all possible $T_{q}$.
However, the random graph intuition (which says that the threshold for
$q$ is {\em asymptotically the same} for the game
played by two ``clever'' players and the game played by two ``random''
players) suggests that Maker already has a winning strategy if $q = (1 - o(1)) \log N$. Beck \cite{BeckTTT}  included the following open problem in his list of the \emph{seven most humiliating open problems}. 

\begin{opprob} \label{opprob:VersionwithMakerTournament}
Is it true that Maker has a winning strategy for the tournament game with $q = (1 - o(1)) \log N$ ?
\end{opprob}
\noindent
We prove that the answer to Open Problem \ref{opprob:VersionwithMakerTournament} is yes.

\begin{theo} \label{theo:Makerbuildtournament}
Maker has a winning strategy for the tournament game with  $q = (1 - o(1)) \log N$ 
\end{theo}

\paragraph{Maker's Strategy}

For proving the claimed results we will analyze adapted versions of the following, natural Maker's Strategy for the ordinary (1:1) $q$-clique game: Maker first selects an arbitrary vertex $v_{1}$. In each of his next moves he claims an edge incident to $v_{1}$ until all edges incident to $v_{1}$ have been occupied. We refer to this sequence of moves as \emph{processing $v_{1}$}. In this way Maker can connect (in his graph) at least $\frac{N - 1}{2}$ vertices to $v_{1}$. So his task is now reduced to achieving a $(q-1)$-clique in a graph with roughly $\frac{N}{2}$ vertices. It seems very plausible that by applying this strategy recursively Maker can, for $q = \log N$, achieve a $q$-clique in $2N$ steps. 
However, there is one obstacle: While Maker claims edges incident to $v_{1}$ Breaker can already claim other edges in the graph, which might later bring Maker into troubles. To prevent this, Maker roughly proceeds as follows. After connecting $v_{1}$ to as many vertices as possible he deletes all vertices whose degree is above some threshold $t$ and then continues in the resulting graph, where every vertex is connected to almost all (i.e., all but at most $t$) of the other vertices. 
By choosing $t$ appropriately Maker can guarantee that both of these restrictions (i.e., fewer vertices and smaller degrees) do not have a significant influence. 
By a careful analysis we can show that Maker can achieve a $(\log N - o(1))$-clique in $N$ steps.
We denote the above strategy by $S$.

For the clique size our result is weaker than Beck's result by a factor of 2. However, the described strategy turns out to be helpful for some variations of the clique game: For the biased clique game we consider the following adaptation of Maker's strategy  $S$: At the beginning, instead of selecting \emph{one} vertex $v_{1}$, he occupies an $m$-clique $C$ in his graph. (In the more detailed analysis in Section \ref{sec:BiasedGame} we will show how Maker can achieve this.) Let $v_{1}, \ldots, v_{m}$ denote the vertices of $C$. 
As long as there are vertices $v$ for which $(v, v_{1}), (v, v_{2}), \ldots, (v, v_{m})$ are all unclaimed, as his move, Maker fixes such a $v$ and connects $v$ to $v_{1}, \ldots, v_{m}$. In this way Maker can achieve that in his graph roughly $\frac{N - m}{(b + 1)}$ vertices are adjacent to \emph{every} $v_{i} \in \{v_{1}, \ldots, v_{m}\}$.

A handwaving analysis (neglecting the fact that Breaker might have claimed edges which are non-incident to a clique edge) gives that Maker can achieve a $K_{q}$ for $q = \frac{m}{\log (b + 1)} \log N$, which is actually roughly the same as we get in our careful analysis.

An adaption of the strategy $S$ can also be used to prove Theorem \ref{theo:Makerbuildsfastclique}: Basically the only additional requirement is that after processing $q$ vertices the required set of $r$ vertices is still present, which causes an additional factor of roughly $r$.

Finally, for the tournament game Maker can adapt his strategy $S$ as follows. Let $T$ be the goal-tournament of Maker on the vertex set 
$\{u_{1}, \ldots, u_{q}\}$. During the game Maker will maintain so called \emph{candidate sets} $V_{1}, \ldots, V_{q}$ such that every $v_{i} \in V_{i}$ is still suitable for the part of vertex $u_{i}$. In round $i$ Maker basically selects a vertex $v_{i}$ in $V_{i}$ and proceeds in such a way that for every $j > i$ he finally possesses $\frac{|V_{j}|}{2}$ edges of the form $(v_{i}, v_{j})$ where $v_{j} \in V_{j}$ and the orientation of $(v_{i}, v_{j})$ equals the orientation of $(u_{i}, u_{j})$. 
In this way Maker reduced his task to occupying a fixed tournament on $q-1$ vertices in the subgraph induced by those vertices in $V_{i + 1} \cup V_{i + 2} \cup \ldots \cup V_{q}$ which are in Maker's graph adjacent to $v_{i}$. Note that in each such round the number of vertices in $V_{j}$ is roughly halved, which suggests that Maker has a winning strategy for $q = (1 - o(1)) \log (N)$.

\paragraph{Notation} Let $G$ be a graph on $n$ vertices and let $v$ be a vertex in $V(G)$. 
By $d(v)$ we denote the ordinary degree of $v$ in $G$.
The \emph{complementary degree} $\bar{d}_{G}(v)$ of $v$ is the number of vertices different from $v$ in $G$ which are non-adjacent to $v$, i.e., $\bar{d}_{G}(v) = n - 1 - d(v)$.
If there is no danger of confusion we sometimes just write ${\bar{d}}(v)$ for ${\bar{d}}_{G}(v)$.

If we consider the course of a game then $d_{B}(v)$ denotes the degree of $v$ in Breaker's graph.
%The \emph{neutral subgraph} is the subgraph of $G$ containing exactly the free edges.

For a subset $S = \{v_{1}, \ldots, v_{i}\} \subseteq V(G)$, the \emph{subgraph induced by $S$}, $G_{[v_{1}, \ldots, v_{i}]}$, denotes the graph obtained from deleting all vertices of $V(G) \backslash S$ in $G$.

%%%%%%%%%%%%%%%%%%%%%%%%%%%%%%%%%%%%%%%%%%%%%%%%%%%%%%%%%%%%%%%%%%

\section{The Biased Game} \label{sec:BiasedGame}

The following is a well known fact in graph theory.

\begin{obse} \label{obse:boundforindepset}
Let $G$ be a graph on $n$ vertices with $\bar{d}(v) \leq d$ for every vertex $v \in V(G)$. Then $G$ contains a clique of size $\frac{n}{d + 1}$.
\end{obse}
\noindent
This can be seen by considering the following greedy algorithm for building a clique: In every round select an arbitrary vertex, add it to the clique and delete all its neighbors. In this way, we deleted at most $d$ vertices per clique-vertex and thus get a clique of size at least $\frac{n}{d + 1}$.

\begin{prop} \label{prop:alwaysachievecliqueforlargegraphs}
For every $q,m,b$ there is an $n = n(q,m,b)$ such that in the $(m:b)$ clique game played on $K_{n}$ Maker has a strategy to achieve a $K_{q}$.
\end{prop}
\noindent
\emph{Proof:} It suffices to consider the case where $m = 1$. 
We proceed by induction on $q$. Clearly, Maker can always achieve a $K_{1}$.
Suppose now that $q > 1$.
Let $n := [5b^{2}(b+1)^{2}] \cdot n(q-1,1,b) + 1$ and let $\{v_{1}, \ldots, v_{n}\}$ be the vertex set of a $K_{n}$. 
Maker uses the following strategy. Until all edges incident to $v_{1}$ have been occupied he claims in each of his moves one edge of the form ($v_{1}, v_{i})$ for some $i$. In this way he can in total occupy at least $\frac{n-1}{b + 1}$ edges incident to $v_{1}$. In the meantime Breaker has claimed at most 
$b \cdot (n - 1)$ edges. Maker iteratively removes every vertex $v \in \{v_{2}, \ldots, v_{n}\}$ with $d_{B}(v) \geq 2b(b+1)$. Let $W$ denote the set of remaining vertices which are in Maker's graph adjacent to $v_{1}$.
By construction, $|W| \geq \frac{n-1}{b + 1} - \frac{b(n - 1)}{2b(b+1)} \geq \frac{n - 1}{2(b+1)}$ and $\bar{d}(v) \leq 2b(b+1)$ for every vertex $v \in W$.
By Observation \ref{obse:boundforindepset}, $W$ contains a clique $K$ of size $n' := \frac{|W|}{2b(b+1) + 1}  \geq \dfrac{\frac{n-1}{2(b+1)}}{2b(b+1) + 1}$. Note that no edge of $K$ has been claimed by either of the players. By our choice of $n$ we have $n' \geq n(q-1,1,b)$ and therefore Maker can achieve a $K_{q-1}$ on $K$, which together with $v_{1}$ forms a $K_{q}$. \hfill $\qed$

\paragraph{Proof of Theorem \ref{theo:Makerbuildscliquebiased}}
Choose $C = C(m,b)$ in such a way that in the $(m:b)$ clique game played on $K_{C}$ Maker has a strategy to achieve a $K_{m}$. (Proposition \ref{prop:alwaysachievecliqueforlargegraphs} guarantees that such a $C$ exists). Note that since we consider $b$ and$ m$ as constants, $C$ is also a constant.
Throughout this section \emph{game} means the $(m:b)$ clique game. 

The next lemma shows how Maker can reduce his task to occupying a clique with $m$ vertices less (than the original clique) in some appropriate subgraph.

\begin{lemm} \label{lemm:Makerrecurrencetosmallerset}
Let $G$ be a graph on $n$ vertices such that $\bar{d}(v) \leq d$ for every $v \in V(G)$ and $n \geq C(d + 1)$. Let $q \geq 1$ and let 
\begin{displaymath}
n' := \frac{n - C - m d - 2b \cdot \binom{C}{2}}{b + 1} - \frac{bn}{q} 
= 
\frac{n}{(b + 1) + \frac{b(b+1)^{2}}{q - b(b+1)}} - \frac{C + m d + 2b \cdot \binom{C}{2}}{b + 1}
\end{displaymath}
Maker can achieve that for some $\{v_{1}, \ldots, v_{m}\} \cup \{w_{1}, \ldots, w_{n'}\} \subseteq V(G)$, (i) every edge $(v_{i}, u)$ with $i \in \{1, \ldots, m\}$ and $u \in \{v_{1}, \ldots, v_{i-1}, v_{i+1}, \ldots, v_{m}\} \cup \{w_{1}, \ldots, w_{n'}\}$ belongs to Maker's graph, (ii) $\bar{d}(w_{i}) \leq d + q$ for every $i$ with $1 \leq i \leq n'$, and (iii) the subgraph induced by $\{w_{1}, \ldots, w_{n'}\}$ contains no Breaker's edge.
\end{lemm}
\noindent
Before proving Lemma \ref{lemm:Makerrecurrencetosmallerset} we first show its consequences.
For integers $n,d$ let $K(n,d)$ denote the class of graphs $G$ on $n$ vertices with $\bar{d}(v) \leq d$ for every $v \in V(G)$. 

\begin{coro} \label{coro:Makerbuildcasespecifically}
Let $d, n, q, s,$ be integers and let $n'$ be defined as in Lemma \ref{lemm:Makerrecurrencetosmallerset}.  If for every $G' \in K(n',d + q)$ Maker can obtain a $K_{s}$ in the game on $G'$ then he can achieve a $K_{s + m}$ in the game on $G$ for every $G \in K(n,d)$.
\end{coro}
\noindent
\emph{Proof of Lemma \ref{lemm:Makerrecurrencetosmallerset}:}
Maker proceeds as follows.
\begin{itemize} 
\item[\textbf{Round 1}] Maker selects a set $S$ of $C$ vertices which form a clique in $G$. (Such a set $S$ exists due to Observation \ref{obse:boundforindepset} and the assumption that $n \geq C(d + 1)$.) Then he occupies a clique $K_{m}$ on $S$ (this is possible by the definition of $C$). Let $v_{1}, \ldots, v_{m}$ denote the clique-vertices. Note that in the meantime Breaker occupied at most $b \binom{C}{2}$ edges.

\item[\textbf{Round 2}] Let $U := V(G) \backslash S$. Maker removes all vertices in $U$ which are incident to at least one Breaker-edge. Let $U'$ denote the resulting vertex set. Note that $|U'| \geq |U| - 2b\binom{C}{2} = n - C - 2b\binom{C}{2}$. Note that the graph induced by $\{v_{1}, \ldots, v_{m}\} \cup U'$ contains no Breaker-edge. However, it is possible that some vertices $u$ in $U'$ are not connected to all vertices in $\{v_{1}, \ldots, v_{m}\}$. (Note that $\bar{d}(u)$ can be larger than 0).
Let $U''$ be the vertex set obtained by deleting all vertices in $U'$ which are non-adjacent to at least one of the $v_{i}$.
Note that 
\begin{equation} \label{eq:resultingvertexsetU}
|U''| \geq |U'| - md \geq |U| - 2b\binom{C}{2} - md = n - C - 2b\binom{C}{2} - md
\end{equation}
 and that every edge $(v_{i}, u)$ with $1 \leq i \leq m$ and $u \in U''$ is present.

\item[\textbf{Round 3}] As long as there are vertices $u \in U''$ where $(u, v_{i})$ is unclaimed for every $i \in \{1, \ldots, m\}$, as his move Maker selects such a $u$ and occupies the edges $(u, v_{1}), (u, v_{2}), \ldots, (u, v_{m})$. Note that he can do at least 
\begin{equation}  \label{eq:vertexsetforend}
n_{\remain} := \frac{|U''|}{b + 1}
\end{equation}
such moves. Let $u_{1}, \ldots,  u_{n_{\remain}}$ denote the corresponding vertices of $U''$. Note that Maker possesses every edge $(u,v_{i})$ with $u \in \{u_{1}, \ldots, u_{n_{\remain}}\}$ and $1 \leq i \leq m$.

\item[\textbf{Round 4}] During Round 3 Breaker has claimed at most $bn$ edges. Maker iteratively deletes every vertex in $\{u_{1}, \ldots, u_{n_{\remain}}\}$ which has degree at least $q$ in Breaker's graph. In this way at most $\frac{bn}{q}$ vertices are deleted (otherwise we would have deleted more edges than Breaker occupied.) Let $\{w_{1}, \ldots, w_{n_{\remain} - \frac{bn}{q}}\} \subseteq \{u_{1}, \ldots, u_{n_{\remain}}\}$ denote the set of non-deleted vertices. Removing all Breaker's edges gives a subgraph with vertex set $\{v_{1}, \ldots, v_{m}\} \cup \{w_{1}, \ldots, w_{n_{\remain} - \frac{bn}{q}}\}$ such that for every $w \in \{w_{1}, \ldots, w_{n_{\remain} - \frac{bn}{q}}\}$, $\bar{d}(w) \leq d + q$, and Maker possesses every edge $(v_{i}, u)$ with $1 \leq i \leq m$ and $u \in \{v_{1}, \ldots, v_{i-1}, v_{i + 1}, \ldots, v_{m}\} \cup \{w_{1}, \ldots  w_{n_{\remain} - \frac{bn}{q}}\}$.

\end{itemize}

We have 
\begin{eqnarray*}
n_{\remain} - \frac{bn}{q} & = & \frac{|U''|}{b + 1} - \frac{bn}{q} \enspace \text{(by \eqref{eq:vertexsetforend})} \\
         & \geq & \frac{n - C - 2b\binom{C}{2} - md}{b + 1} - \frac{bn}{q} \enspace \text{(by  \eqref{eq:resultingvertexsetU})} \\
       & = & n'
\end{eqnarray*}
\hfill $\qed$

We can analyze Maker's strategy by applying Corollary \ref{coro:Makerbuildcasespecifically} repeatedly.
Let 
\begin{equation} \label{eq:showthatMakersstrategysuccessfulsetq}
q := \left(\frac{m}{\log(b + 1)}\right) \cdot (\log N - 5\log\log N)
\end{equation}
For simplicity we assume that $q$ is divisible by $m$. (For the case where $q$ is not divisible by $m$ we can then follow similar lines.)
Our goal is to show that in the game on $K_{N}$ Maker can achieve a $K_{q}$.

Let $n$ be a large enough integer and let $n'$ be defined as in Lemma \ref{lemm:Makerrecurrencetosmallerset}. Then
\begin{equation} \label{eq:simplifyexpressionnumbervertices}
n' \geq \frac{n}{(b + 1) + \frac{b(b+1)^{2}}{q - b(b+1)}} - (c_{1}d + c_{2})  
\end{equation} 
for appropriate constants $c_{1}, c_{2} \geq 0$.

\begin{prop} \label{prop:Makerwinsuseinduction}
Let $r := b + 1 + \frac{b(b+1)}{q - b(b+1)}$. Let $i \in \{1, \ldots, \frac{q}{m}\}$. If 
\begin{displaymath}
\frac{n}{r^{i}} - i \cdot (c_{1}q^{2} + c_{2}) \geq C(q^{2} + 1)
\end{displaymath}
then for every $G \in K(n, (\frac{q}{m} - i) \cdot q)$,
Maker can achieve a $K_{i \cdot m}$ in the game on $G$.
\end{prop}
\noindent 
\emph{Proof:} We apply induction. For $i = 1$ the claim is clearly true. Indeed, if $\frac{n}{r} - (c_{1}q^{2} + c_{2}) \geq C(q^{2} + 1)$ then $n \geq C(q^{2} + 1)$. By assumption and Observation \ref{obse:boundforindepset} $G$ contains a clique of size at least $\frac{n}{q^{2} + 1} \geq C$. By our choice of $C$ Maker can obtain the desired clique.

Assume now that $i \geq 2$. Let $G \in K(n, (\frac{q}{m} - i) \cdot q)$ and let $n'$ be defined as in Lemma \ref{lemm:Makerrecurrencetosmallerset}.
Suppose that 
\begin{equation} \label{eq:conditionfoclaimforindusat}
\frac{n}{r^{i}} - i \cdot (c_{1}q^{2} + c_{2}) \geq C(q^{2} + 1)
\end{equation}
By \eqref{eq:simplifyexpressionnumbervertices} (for $d = (\frac{q}{m} - i) \cdot q$) we obtain $n \leq r \cdot (n' + c_{1} \cdot (\frac{q}{m} - i) \cdot q + c_{2}) \leq r \cdot (n' + c_{1} q^{2} + c_{2})$. Together with  \eqref{eq:conditionfoclaimforindusat} this gives 
\begin{equation} \label{eq:proofofindupropforbiased}
\frac{n' + c_{1}q^{2} + c_{2}}{r^{i - 1}} - i \cdot (c_{1}q^{2} + c_{2}) \geq C(q^{2} + 1)
\end{equation}
Thus
%Since $\frac{n' + c_{1} \cdot q^{2} + c_{2}}{r^{i - 1}} \leq \frac{n'}{r^{i - 1}} + c_{1}d + c_{2}$ we obtain the following.
\begin{displaymath}
\frac{n'}{r^{i - 1}} - (i-1) \cdot (c_{1}q^{2} + c_{2}) =
\frac{n'}{r^{i - 1}} + c_{1}q^{2} + c_{2} - i \cdot (c_{1}q^{2} + c_{2}) \geq 
\frac{n' + c_{1}q^{2} + c_{2}}{r^{i - 1}} - i \cdot (c_{1}q^{2} + c_{2}) \geq C(q^{2} + 1)
\end{displaymath}
By induction Maker can achieve a $K_{(i - 1)m}$ in the game on $G'$ for every 
$G' \in K(n, (\frac{q}{m} - (i - 1)) \cdot q)$. Together with Corollary \ref{coro:Makerbuildcasespecifically} this concludes our proof. \hfill $\qed$

We now complete the proof of Theorem \ref{theo:Makerbuildscliquebiased}. Note that $K_{N}$ is the unique element of $K(N,0)$. 
By Proposition \ref{prop:Makerwinsuseinduction} (for $i = \frac{q}{m}$) Maker can achieve a $K_{q}$ in the game on $K_{N}$ if 
\newline
 $\frac{N}{r^{\frac{q}{m}}} - \frac{q}{m} \cdot (c_{1}q^{2} + c_{2}) \geq C(q^{2} + 1)$.
\newline
Recall that $r =  b + 1 + \frac{b(b+1)}{q - b(b+1)}$. We have $r \leq (b + 1) \cdot (1 + \frac{b}{q - b(b+1)}) \leq (b + 1) \cdot e^{\frac{b}{q - b(b+1)}} \leq (b + 1) \cdot e^{\frac{2b}{q}}$. Hence $r^{\frac{q}{m}} \leq (b + 1)^{\frac{q}{m}} \cdot e^{2b}$. Thus
\begin{equation} \label{eq:getrelofNandqforbiasedcliquegame}
\frac{N}{r^{\frac{q}{m}}} \geq 
\frac{N}{(b + 1)^{\frac{q}{m}} \cdot e^{2b}} 
\end{equation}
Hence
\begin{eqnarray*}
\frac{N}{r^{\frac{q}{m}}} - \frac{q}{m} \cdot (c_{1}q^{2} + c_{2}) & \geq & \frac{N}{r^{\frac{q}{m}}} - q^{4} \\
& \geq & 
\frac{N}{(b + 1)^{\frac{q}{m}} \cdot e^{2b}}  - q^{4} \enspace \text{(by \eqref{eq:getrelofNandqforbiasedcliquegame})}\\
& \geq &
\frac{N}{(b + 1)^\frac{\log N - 5\log\log N}{\log(b + 1)} \cdot e^{2b}} - \left(\frac{m}{\log(b + 1)}\right)^{4} \log^{4}N \enspace \text{(by \eqref{eq:showthatMakersstrategysuccessfulsetq})} \\
& \geq & \frac{N}{2^{\log N - 5\log\log N} \cdot e^{2b}} - \left(\frac{m}{\log(b + 1)}\right)^{4}  \log^{4}N \\
& \geq & \frac{\log^{5}N}{e^{2b}} - \left(\frac{m}{\log(b + 1)}\right)^{4}  \log^{4}N \\
& \geq & C \cdot \left(\left(\frac{m}{\log(b + 1)}\right)^{2}  \log^{2}N + 1 \right) \\
& \geq & C(q^{2} + 1)
\end{eqnarray*}
By Proposition \ref{prop:Makerwinsuseinduction} Maker can achieve the required clique. \hfill $\qed$

\section{Building a Clique Fast} \label{sec:CliqueFast}

Throughout this section by \emph{game} we mean the ordinary (1:1) clique game. 
For integers $n,d$ let $K(n,d)$ denote the class of graphs $G$ on $n$ vertices with $\bar{d}(v) \leq d$ for every $v \in V(G)$. 

\emph{Proof of Theorem \ref{theo:Makerbuildsfastclique}:} 
Let $C(q,r)$ denote the constellation Maker is claimed to achieve in Theorem \ref{theo:Makerbuildsfastclique}.

\begin{lemm} \label{lemm:Makerrecurrenceforfastclique}
Let $G \in K(n,d)$, let $q \geq 1$ and let $v_{1} \in V(G)$.
Maker can achieve in $\frac{n}{2}$ moves that for some $W \subseteq V(G) \backslash \{v_{1}\}$ with $|W| = \frac{n - 1 - d}{2} - \frac{n}{q}$,  (i)  $(v_{1},w)$ belongs to Maker's graph for every $w \in W$, (ii) $\bar{d}(w) \leq d + q$ for every $w \in W$, and (iii) the subgraph induced by $W$ contains no Breaker's edge.
\end{lemm}
\noindent
\emph{Proof:} 
Maker proceeds as follows.
\begin{itemize} 
\item[\textbf{Round 1}] Maker removes all vertices in $V(G) \backslash \{v_{1}\}$ which are non-adjacent to $v_{1}$. Note that by assumption at most $d$ vertices are deleted.
In his next $\frac{n - 1 - d}{2}$ moves Maker occupies an unclaimed edge incident to $v_{1}$. (Since there are at least $n - 1 - d$ vertices and in the first $\frac{n - 1 - d}{2}$ moves Breaker can collect at most $\frac{n - 1 - d}{2}$ edges, Maker can make these moves.)
Let $V'$ denote the set of vertices which are in Maker's graph adjacent to $v_{1}$. Note that 
$|V'| \geq \frac{n - 1 - d}{2}$.

\item[\textbf{Round 2}] During Round 1 Breaker has occupied at most $n$ edges. Let $W$ denote the vertex set resulting from deleting iteratively every vertex $v$ with Breaker's degree at least $q$ from $V'$. Note that $|W| \geq |V'| - \frac{n}{q}$ (otherwise the number of Breaker's edges which were deleted is larger than $n$). Finally we delete all Breaker's edges in $W$, which increases $\bar{d}(w)$ by at most $q$ for every $w \in W$. Clearly, $v_{1}$ and $W$ fulfill (i), (ii) and (iii).
\end{itemize}
Note that Maker makes at most $\frac{n - 1 - d}{2} \leq \frac{n}{2}$ moves during Round 1 and no move during Round 2. So the required constellation can be obtained in $\frac{n}{2}$ moves.
%lasts $\frac{n - 1 - d}{2} \leq \frac{n}{2}$ Breaker's moves and Step 2 does not contain a move
\hfill $\qed$

We have
\begin{equation} \label{eq:alternativeformulationofsmallern}
\frac{n - 1 - d}{2} - \frac{n}{q} = \frac{n}{2 + \frac{4}{q - 2}} - \frac{d + 1}{2} \geq \frac{n}{2 + \frac{4}{q - 2}} - (d + 1)
\end{equation}
The following is a consequence of Lemma \ref{lemm:Makerrecurrenceforfastclique} and \eqref{eq:alternativeformulationofsmallern}.
\begin{coro} \label{coro:Makerbuildaddverttoclique}
Let $d, i, n, q, r, s$ be integers.  
If for every $G' \in K\left(\frac{n}{2 + \frac{4}{q-2}} - (d + 1), d + q \right)$ Maker can in $s$ moves
obtain a $C(i-1, r)$ on $G'$ then he can achieve a $C(i,r)$  on $G$ in $s + \frac{n}{2}$ moves for every $G \in K(n,d)$.
\end{coro}
\noindent
We will analyze Maker's strategy by applying Corollary \ref{coro:Makerbuildaddverttoclique} repeatedly.
We fix a $q$ and let
\begin{equation} \label{eq:forfastcliquedefN}
N := q^{5} \cdot 2^{q} \cdot r
\end{equation}

Our goal is to show that in the game on $K_{N}$ Maker can achieve a $C(q,r)$.

\begin{prop} \label{prop:Iterativelyuseprevcoro}
Let $i \in \{0, 1, \ldots, q\}$.
If 
\begin{displaymath}
\frac{n}{\left(2 + \frac{4}{q-2}\right)^{i}} - i q^{2} \geq r(q^{2} + 1)
\end{displaymath}
then for every $G \in K(n, (q - i) \cdot q)$,
Maker can achieve a $C(i,r)$ on $G$ in $n$ moves.
\end{prop}
\noindent 
\emph{Proof:} We proceed by induction. 
We first consider the case where $i = 0$. Let $G \in K(n, q^{2})$ and suppose that
$n \geq r(q^{2} + 1)$.
%Maker first removes every vertex of $V(G) \backslash \{v_{1}\}$ from $G$ (so he deletes at most $q^{2}$ vertices) and then claims an unclaimed edge incident to $v_{1}$ in each of his next $\frac{n - 1 - q^{2}}{2}$ moves. (He can do this since Breaker collected at most $\frac{n - 1 - q^{2}}{2}$ incident to $v_{1}$ in his first $\frac{n - 1 - q^{2}}{2}$.) Let $V' \subseteq V(G) \backslash \{v_{1}\}$ denote the subset containing the vertices which are in Maker's graph adjacent to $v_{1}$. Note that $|V'| \geq \frac{n - 1 - q^{2}}{2}$.
By Observation \ref{obse:boundforindepset} $G$ contains a clique of size $\frac{n}{q^{2} + 1}$ which by assumption is at least $r$. So there is a $C(0, r)$ in $G$.

Suppose now that $i \geq 1$ and assume that
\begin{equation} \label{eq:determinefastcliquerecursiononeback}
\frac{n}{\left(2 + \frac{4}{q-2}\right)^{i}} - i q^{2} \geq r(q^{2} + 1)
\end{equation}
Let 
\begin{equation} \label{eq:reformulationofnumvertforindu}
n':= \frac{n}{2 + \frac{4}{q - 2}} - ((q-i) \cdot q + 1) \geq \frac{n}{2 + \frac{4}{q - 2}} - q^{2}
\end{equation}
Note that $n \leq (n' + q^{2}) \cdot \left(2 + \frac{4}{q - 2}\right)$. By \eqref{eq:determinefastcliquerecursiononeback}, 
\begin{displaymath}
r(q^{2} + 1) \leq \frac{n' + q^{2}}{\left(2 + \frac{4}{q-2}\right)^{i - 1}} - i q^{2} \leq 
\frac{n'}{\left(2 + \frac{4}{q-2}\right)^{i - 1}} + q^{2} - i q^{2} = 
\frac{n'}{\left(2 + \frac{4}{q-2}\right)^{i - 1}} - (i-1)q^{2}
\end{displaymath}
By induction, for every $G \in K(n', (q - i + 1) \cdot q)$
Maker can achieve a $C(i - 1,r)$ on $G$ in $n'$ moves. 
By Corollary \ref{coro:Makerbuildaddverttoclique} for $d = (q - i) \cdot q$ and $s = n'$, Maker can achieve a $C(i, r)$ on $G$ in $\frac{n}{2} + n'$ moves, for every $G \in K(n, q - i)$. Since $n' \leq \frac{n}{2}$ the number of moves is at most $n$. \hfill $\qed$

We now conclude the proof of Theorem \ref{theo:Makerbuildsfastclique}: Note that 
$K_{N}$ is the unique element of $K(N,0)$. It suffices to show that Maker can achieve a $C(q,r)$ on $K_{N}$ in $N$ moves.
We have
 \begin{equation} \label{eq:actuallydivderoughlybytwo}
\left(2 + \frac{4}{q-2}\right)^{q} = 2^{q} \cdot \left(1 + \frac{2}{q-2} \right)^{q} \leq 2^{q} \cdot \left(1 + \frac{4}{q}\right)^{q} \leq 2^{q} \cdot e^{4}
\end{equation}
Hence by  \eqref{eq:forfastcliquedefN} and \eqref{eq:actuallydivderoughlybytwo},
\begin{displaymath}
\frac{N}{\left(2 + \frac{4}{q-2}\right)^{q}} - q^{3}  \geq 
\frac{N}{\left(2^{q} \cdot e^{4} \right)} - q^{3} \geq 
\frac{q^{5} \cdot r}{e^{4}} - q^{3} \geq r(q^{2} + 1)
\end{displaymath}
Proposition \ref{prop:Iterativelyuseprevcoro} for $i := q$ and $n := N$ yields that on $K_{N}$ Maker can achieve a $C(q,r)$ in $N = q^{5} \cdot 2^{q} \cdot r$ moves. \hfill $\qed$

\section{Building a Tournament} \label{sec:BuildTournament}

We need some notation first.
We assume that Maker colors his edges red and Breaker colors his edges blue.
We say that Maker \emph{wins $T_{s}$ on $G$} if for every tournament $T$ on $s$ vertices Maker has a strategy to achieve a red copy of $T$ in the (1:1) game on $G$.

\emph{Proof of Theorem \ref{theo:Makerbuildtournament}:}

The next lemma describes how Maker can reduce his task of occupying a fixed tournament $T$ to the task of occupying a given tournament with one vertex less. In addition to the clique game Maker will maintain so called \emph{candidate sets} $V_{1}, \ldots, V_{s}$ in such a way that every vertex $v_{i} \in V_{i}$ is still suitable for the part of vertex $i$.

\begin{lemm} \label{lemm:LetTournamentbecomesmallerwithedges}
Let $G$ be a graph such that $\bar{d}(v) \leq d$ for every $v \in V(G)$.
Let $q, r$ be integers, let $T_{r}$ be a fixed tournament on the vertices $u_{1}, \ldots, u_{r}$ and let $V_{1}\uplus V_{2} \uplus \ldots \uplus V_{r}$ be a partition of $V(G)$ such that $|V_{i}| \geq n$ for every $i$ with $1 \leq i \leq r$. 
Maker can achieve that for some $v_{1} \in V_{1}$ and for some subsets $V'_{2}, V'_{3}, \ldots, V'_{r}$ with $V'_{i} \subseteq V_{i}$ for $i$, $2 \leq i \leq r$, (i) for every $i$ with $2 \leq i \leq r$, $v_{1}$ is in Maker's graph adjacent to at least $\frac{n - d}{2} - \frac{rn}{q^{2}}$ vertices $v_{i}$ in $V'_{i}$ in such a way that the orientation of $(v_{1}, v_{i})$ equals the orientation of $(u_{1}, u_{i})$, (ii) $\bar{d}(v) \leq d + q^{2}$, for every $v \in V'_{2} \cup V'_{3} \cup \ldots \cup V'_{r}$, and (iii) the subgraph induced by $V'_{2} \cup V'_{3} \cup \ldots \cup V'_{r}$ contains no Breaker's edge.
\end{lemm}
\noindent
\emph{Proof:} We can assume wlog that $|V_{i}| = n$ for every $i$.
Maker selects an arbitrary vertex $v_{1} \in V_{1}$. Then he proceeds as follows.
\begin{itemize}
\item[\textbf{Round 1}] Maker removes all vertices in $V(G) \backslash \{v_{1}\}$ which are non-adjacent to $v_{1}$. For every $i$, let $\widetilde{V_{i}}$ denote the set of vertices in $V_{i}$ which were not deleted. Note that $|\widetilde{V_{i}}| \geq |V_{i}| - d$. 

\item[\textbf{Round 2}]
Until all edges incident to $v_{1}$ have been claimed Maker applies the following strategy: If Breaker occupies an edge $(v_{1}, v_{i})$ with $v_{i} \in \widetilde{V_{i}}$ for some $i \in \{1, \ldots, r\}$ then Maker occupies -- if possible -- an edge $(v_{1}, w_{i})$ with $w_{i} \in \widetilde{V_{i}}$. If Maker cannot occupy such an edge (because all edges connecting $v_{1}$ with $\widetilde{V_{i}}$ are already occupied) or if Breaker occupies an edge which is not incident to $v_{1}$ then Maker claims an arbitrary edge incident to $v_{1}$. In this way Maker
can in $rn$ moves achieve to possess 
$\frac{|\widetilde{V_{i}}|}{2} \geq \frac{|V_{i}| - d}{2} \geq \frac{n - d}{2}$ edges connecting $v_{1}$ with $\widetilde{V_{i}}$ for every $i$ with $2 \leq i \leq r$. Hence for $i$, $i = 2, \ldots, r$ there is a subset $W_{i} \subseteq V_{i}$ with  
\begin{equation} \label{eq:tournamentMakerfirstselection}
|W_{i}| \geq \frac{n - d}{2}
\end{equation} 
such that Makers graph contains $(v_{1}, w)$ for every $w \in W_{i}$. 

\item[\textbf{Round 3}] 
In Round 2 
%In the meantime
Breaker occupied at most $rn$ edges. In every $W_{i}$ Maker iteratively removes those vertices with Breaker's degree at least $q^{2}$. Let $V'_{i} \subseteq W_{i}$ denote the vertices in $W_{i}$ which were not deleted. Note that $|V'_{i}| \geq |W_{i}| - \frac{rn}{q^{2}}$ (otherwise Maker deleted more edges than Breaker added). 
\end{itemize}

Hence  $|V'_{i}| \geq |W_{i}| - \frac{rn}{q^{2}} \geq \frac{n - d}{2} - \frac{rn}{q^{2}} $ (the last inequality is by \eqref{eq:tournamentMakerfirstselection}). Deleting all Breaker's edges increases $\bar{d}(v)$ by at most $q^{2}$ for every $v \in V'_{1} \cup V'_{2} \cup \ldots, \cup V'_{r}$ and assures (iii). \hfill $\qed$

For integers $d,n,r$ let $K(n,r,d)$ denote the class of graphs $G$ with a partition $V_{1} \uplus V_{2} \uplus \ldots \uplus V_{r}$ of $V(G)$ such that $|V_{i}| \geq n$ for every $i \in \{1, \ldots, r\}$ and $\bar{d}(v) \leq d$ for every $v \in V(G)$.
Lemma \ref{lemm:LetTournamentbecomesmallerwithedges} directly implies the following.

\begin{coro} \label{coro:Makerbuildcasesforpoint}
Let $d, n, r, s$ be integers with $r \geq 2$.  
\newline
If for every $G' \in K(\frac{n - d}{2} - \frac{rn}{q^{2}}, r - 1, d + q^{2})$ Maker can win $T_{s}$ on $G'$ then he can win $T_{s + 1}$ on $G$ for every $G \in  K(n, r, d)$.
\end{coro}
\noindent

We can analyze Maker's strategy by applying Corollary \ref{coro:Makerbuildcasesforpoint} repeatedly.
Let 
\begin{equation} \label{eq:determineqfortournamentgamecloselogN}
q :=  \log N - 6 \log\log N
\end{equation}

Our goal is to show that Maker can win $T_{q}$ on $K_{N}$.

\begin{prop} \label{prop:ConditionforMakerswinondegvert}
Let $i \in \{1, \ldots, q\}$.
If
\begin{displaymath}
\frac{n}{(2 + \frac{4}{q - 2})^{i}} - i \cdot q^{3} \geq 1
\end{displaymath}
then for every $G \in K(n, i, (q- i) q^{2})$ Maker wins $T_{i}$ on $G$.
\end{prop}
\noindent 
\emph{Proof:} We apply induction. For $i = 1$ the claim is clearly true. Indeed, if 
$\frac{n}{2 + \frac{4}{q - 2}} - q^{3} \geq 1$ then $n \geq (q-1)q^{2}$ and therefore $K(n, i, (q - 1) \cdot q^{2})$ is well-defined. Since Maker wins $T_{1}$ on every graph we are done for the case where $i = 1$. 

Suppose now that $i \geq 2$.
Let  $G \in K(n, i, (q- i) q^{2})$ and assume that 
\begin{equation} \label{eq:ProceedwithMakerobtaininductivebound}
\frac{n}{(2 + \frac{4}{q - 2})^{i}} - i \cdot q^{3} \geq 1
\end{equation}
Let
\begin{displaymath} 
n' := \frac{n - (q-i)q^{2}}{2} - \frac{in}{q^{2}} \geq \frac{n - (q-i)q^{2}}{2} - \frac{n}{q} = 
\frac{n}{2 + \frac{4}{q - 2}} - \frac{(q-1)q^{2}}{2}
\end{displaymath}
Note that  $n \leq (n' + \frac{(q-1)q^{2}}{2})(2 + \frac{4}{q-2})$. Together with \eqref{eq:ProceedwithMakerobtaininductivebound} this implies the following.
\begin{equation} \label{eq:GetTerminnandsmallernforrecu}
\frac{n' + \frac{(q-1)q^{2}}{2}}{(2 + \frac{4}{q - 2})^{i - 1}} - iq^{3} \geq 1
\end{equation}
We have 
\begin{displaymath}
1 \leq \frac{n' + \frac{(q-1)q^{2}}{2}}{(2 + \frac{4}{q - 2})^{i - 1}} - iq^{3} \leq 
\frac{n'}{(2 + \frac{4}{q - 2})^{i - 1}} + q^{3} - iq^{3} =
\frac{n'}{(2 + \frac{4}{q - 2})^{i - 1}} - (i-1)q^{3}
\end{displaymath}
By induction Maker can win $T_{i-1}$ on $G'$ for every $G' \in K(n', i - 1, (q - i + 1) q^{2})$. Together with 
Corollary \ref{coro:Makerbuildcasesforpoint} this implies that Maker wins $T_{i}$ on $G$. \hfill $\qed$

We now complete the proof of Theorem \ref{theo:Makerbuildtournament}. We just need to show that Maker can win $T_{q}$ on all $G \in K(\frac{N}{q}, q, 0)$. (We assume here for simplicity that $N$ is divisible by $q$.)
By Proposition \ref{prop:ConditionforMakerswinondegvert} (for $i = q$) we just need to show that 
$\frac{\frac{N}{q}}{(2 + \frac{4}{q - 2})^{q}} -  q^{4} \geq 1$. 
\newline
Note that 
\begin{equation} \label{eq:finalcalculationuseProp}
\left(2 + \frac{4}{q - 2}\right)^{q} = 2^{q} \cdot \left(1 + \frac{2}{q-2}\right)^{q} \leq 2^{q} \cdot e^{\frac{2q}{q-2}} \leq 2^{q} \cdot e^{4}
\end{equation}
By \eqref{eq:determineqfortournamentgamecloselogN} and \eqref{eq:finalcalculationuseProp} we have
\begin{displaymath}
\frac{\frac{N}{q}}{(2 + \frac{4}{q - 2})^{q}} -  q^{4} \geq \frac{N}{q \cdot 2^{q} \cdot e^{4}} - q^{4} \geq
\frac{N}{\log N \cdot 2^{\log N - 6 \log\log N} \cdot e^{4}} - \log^{4}N \geq \frac{\log^{5}N}{e^{4}} - \log^{4}N \geq 1
\end{displaymath}
By Proposition \ref{prop:ConditionforMakerswinondegvert} Maker can win $T_{q}$ on all $G \in K(\frac{N}{q}, q, 0)$. Therefore he can win $T_{q}$ on $K_{N}$. \hfill $\qed$

%\bibliography{biblio}
% \bibliographystyle{plain}

\end{document}